\documentclass[final,5p,times,twocolumn]{elsarticle}

\usepackage{amssymb}

\usepackage{lineno}

\usepackage{lmodern}
\usepackage{graphicx}
\usepackage{xcolor}
\usepackage{amsmath}
\usepackage{amssymb}
\usepackage{array}

\journal{arxiv}

\begin{document}

\begin{frontmatter}



\title{Parallel implementations of random time algorithm for chemical network stochastic simulations}


\author[ciac]{Chuanbo Liu}
\author[ciac,stonybrook]{Jin Wang\corref{cor1}}
\cortext[cor1]{To whom correspondence should be addressed. Email: jin.wang.1@stonybrook.edu}
\address[ciac]{State Key Laboratory of Electroanalytical Chemistry, Changchun Institute of Applied Chemistry, Chinese Academy of Sciences, Jilin, People's Republic of China}
\address[stonybrook]{Department of Chemistry, Physics and Applied Mathematics, State University of New York at Stony Brook, Stony Brook, USA}

\begin{abstract}
In this study, we have developed a parallel version of the random time simulation algorithm. 
	Firstly, we gave a rigorous basis of the random time description of the stochastic process of chemical reaction network time evolution. 
	And then we reviewed the random time simulation algorithm and gave the implementations for the parallel version of next reaction random time algorithm. 
	The discussion of computational complexity suggested a factor of $M$ (which is the connection number of the network) folds time consuming reduction for random time simulation algorithm as compared to other exact stochastic simulation algorithms, such as the Gillespie algorithm. 
  For large-scale system, such like the protein-protein interaction network, $M$ is on order of $10^8$. 
  We further demonstrate the power of random time simulation with a GPGPU parallel implementation which achieved roughly 100 folds acceleration as compared with CPU implementations. 
	Therefore the stochastic simulation method we developed here can be of great application value for simulating time evolution process of large-scale network. 
\end{abstract}


\end{frontmatter}


\section{Introduction}
\label{S:1}

Chemical reaction network time evolution is intrinsically stochastic. 
Starting by Gillespie \cite{Gillespie1997}, a lot of algorithms had been developed for stochastic simulation of the time evolution of chemical reaction network, 
such like the First Reaction Method (FRM), Direct Method (DM). 
Some improved methods had also been developed to accelerate the simulation processes. 
Such like the Optimized Direct Method (ODM), computes $\sum a_j$ incrementally.
The Next Reaction Method (NRM), use one single random number for each simulation step. 
Others like Sorting Direct Method (SDM), 
Logarithmic Direct Method (LDM), 
Partial-propensity Stochastic Simulation Algorithm (PSSA), 
PSSA-composition rejection (PSSA-CR), 
and Sorting Partial-propensity Direct Method (SPDM). 
Focused on different aspects of the simulation steps. 

Development of hardware, especially for the General Purpose Graphics Processing Units (GPGPU), give the opportunity for developing parallel algorithm that can take advantage of the large number of thread blocks (TBs). 
Various data structure and implementations were also been developed for acceleration of the stochastic simulation on fine-grid or coarse-grid level. 
These implementations can achieve about 10 folds of simulation acceleration as compared to mature CPU based simulation toolkit, such like StochKit. 
We here demonstrated a new implementation of random time algorithm with GPGPU for stochastic simulations. 
By carefully arranging the data distribution on global and local memories we can accelerate the stochastic simulation with a data parallel manner for roughly 100 folds. 
In this paper, we first briefly reviewed the methods of random time simulation, then discussed the relative computational complexity. 
At last, we demonstrated the methods with an oscillation predator-prey model. 

\section{A brief introduction to the random time approach}
In this part, we follow \cite{Anderson2011}. 
But we are trying to give a rigorous generalized approach.

Consider $N\geqslant1$ chemical species in a well-mixed system. 
Also consider $M\geqslant1$ reactions, labeled with $k$, the stoichiometric number of $j$-th chemical species in $k$-th reaction is $\sigma_{jk}$. 
The simplest stochastic models for describing the system is a continuous time Markov chain.
The state is represented by the molecule number of chemical species $X(t)=\{X_1, X_2, \ldots, X_N\}$ and reactions modeled as possible transitions of the chain. 
It was shown by Gillespie \cite{Gillespie1992, Gillespie1997} that in a well mixed system, 
the probability for a specific reaction takes place is governed by the propensity function $a_k(X(t),t)$ when consider time-inhomogeneous chemical reaction networks. 
The propensity function is
\begin{equation}\label{eq:propensity_function_definition}
a_k(X(t),t)=c_k(t)h_k(X(t))
\end{equation}
where 
\begin{equation}\label{eq:propensity_function_hk}
h_k(X(t))=\prod_{j}\binom{X_k(t)}{\sigma_{jk}}=\prod_j \frac{X_k(t)!}{\sigma_{jk}!(X_k(t)-\sigma_{jk})!}
\end{equation}
when considering the time evolution , and the state vector for $k$-th jump is $v_k$. 
With Markov property, the conditional probability can be expressed as:
\begin{equation}\label{eq:conditional_probability}
	\mathbf{P}\{\text{$k$-th reaction fired once in }(t, t+\Delta t) | \mathcal{F}_t\} = a_k(X(t), t) \Delta t
\end{equation}
where $\mathcal{F}_t$ is the $\sigma$-algebra representing the information about the system that is available at time t \cite{Kolmogorov1956}.
The consequence of Eq. \ref{eq:conditional_probability} is the possibility of two reactions occurred at the same is on the order of $(\Delta t)^2$, which means it is very unlikely to happen. 
This can be shown by simply calculate the joint probability as
\begin{equation}\label{eq:events_occure_at_a_time}
	\begin{aligned}
  & \mathbf{P}\{R_k(t+\Delta t)-R_k(t)\geqslant 2\}\\
  &=\mathbf{P}\{R_k(t + \Delta t) - R_k(t)\}^2 \\
	& \sim \mathbf{P}\{R_k(t+\Delta t)-R_k(t)\geqslant 1, R_j(t+\Delta t)-R_j(t)\geqslant 1\}\\
	&= \mathbf{P}\{R_k(t+\Delta t)-R_k(t)\}\cdot \mathbf{P}\{R_j(t+\Delta t) - R_j(t)\}\\
	&\sim  O((\Delta t)^2)
	\end{aligned}
\end{equation}

The system evolution can be described equivalently as a counting process by replacing the random variables from chemical species molecule number to number of times that a specific reaction fired. 
If $R_k(t)$ is the number of times that the $k$-th reaction has fired up to time $t$, then the state at $t$ that originated from $X(0)$ is given by
\begin{equation}\label{eq:space_system_evolve}
X(t)=X(0)+\sum_{k=1}^M R_k(t) v_k
\end{equation} 
This counting process is a generalized Poisson process $G(\lambda(t), t)$ in the sense that the arrival rate $\lambda$ is a function of time $t$. 
The total arrival times for generalized Poisson process $G$ can be calculated as
\begin{equation}\label{eq:stochastic_integral_G}
	R(t) = \int_0^t dG(\lambda(s), s)
\end{equation}
The Poisson distribution with parameter $\lambda$ in the time interval $(s, t]$ describe the arrival times of events of a Poisson process $N$ in this time interval. 
\begin{equation}\label{eq:Poisson_distribution}
	\mathbf{P}\{N(\lambda, s) - N(\lambda, t) = k\} = \frac{(\lambda (s - t))^k}{k!} e^{-\lambda(s-t)}
\end{equation}
Therefore the probability of $k$ times events happened in the time interval $(t, t+\Delta t]$ is
\begin{equation}\label{eq:G_Y_1}
	\begin{aligned}
	&\mathbf{P}\{R(t + \Delta t) - R(t) = k | \mathcal{F}_t\} \\
	&= \mathbf{P}\{\int_t^{t+\Delta t} dG(\lambda(s), s) = k | \mathcal{F}_t\}  \\ 
	&=\mathbf{P}\{G(\lambda(t), t + \Delta t) - G(\lambda(t), t) = k | \mathcal{F}_t\} \\ 
	&=\frac{(\lambda(t)\Delta t)^k}{k!} e^{-\lambda(t) \Delta t}  \\
	&=\frac{\left(\int_t^{t+\Delta t} \lambda(s)ds\right)^k}{k!} e^{-\int_t^{t+\Delta t}\lambda(s)ds} \\
	&=\mathbf{P}\left\{Y\left(\int_t^{t+\Delta t} \lambda(s)ds\right) - Y\left(\int_0^{t} \lambda(s)ds\right) = k | \mathcal{F}_t\right\} \\
	\end{aligned}
\end{equation}
where $Y$ is unit rate Poisson process.
The trick here is $\lambda$ is not changed in the infinitesimal time interval $(t, t+\Delta t]$.
Also it can be noticed
\begin{equation}\label{eq:trick}
	\lim_{\Delta t \rightarrow 0} \int_t^{t + \Delta t} \lambda(s) ds = \lim_{\Delta t \rightarrow 0} \lambda(t) \Delta t
\end{equation}
With Eq. \ref{eq:Poisson_distribution}, the result is obvious. 
Eq. \ref{eq:G_Y_1} suggests
\begin{equation}\label{eq:R(t)_Y}
	R(t) = Y\left(\int_0^{t} \lambda(s)ds\right)
\end{equation}

Next we can establish the relationship between $\lambda$ and the propensity function $a_k$. 
We can now rewrite the left side of Eq. \ref{eq:conditional_probability} as
\begin{equation}\label{eq:rewrite_conditional_probability_left}
	\begin{aligned}
	&\mathbf{P}\left\{R_k(t+\Delta t) - R_k(t) = 1 | \mathcal{F}_t \right\} \\
	&= \left(\int_t^{t+\Delta t} \lambda_k(s) ds \right) e^{-\left(\lambda_k(s)ds\right)} \\
	&= \left(\int_t^{t+\Delta t} \lambda_k(s)ds\right) + O((\Delta t)^2)  \\
	\end{aligned}
\end{equation}
Compared to the right side of Eq. \ref{eq:conditional_probability}, we can have
\begin{equation}\label{eq:lambda_propensity_prepare}
	\int_t^{t+\Delta t} \lambda_k(s)ds = a_k(X(t), t) \Delta t
\end{equation}
which means, 
\begin{equation}\label{eq:lambda_propensity}
	\lambda_k(s) = a_k(X(t), t)
\end{equation}
Therefore, from Eq. \ref{eq:space_system_evolve}, Eq. \ref{eq:R(t)_Y} and Eq. \ref{eq:lambda_propensity}, the system evolution equation is, 
\begin{equation}\label{eq:system_evolution_equation}
	X(t) = X(0) + \sum_{k=1}^M Y_k\left(\int_0^{t} a_k(X(s), s) ds\right) v_k
\end{equation}

So we have represented the chemical reaction network evolution process as an increment counting process. 
Every infinitesimal time frame of this stochastic process can be further decomposed to $M$ independence unit rate Poisson processes. 
Eq. \ref{eq:system_evolution_equation} can be rewritten to have the same formula with Eq. \ref{eq:space_system_evolve} by introducing the ``internal time'' for each chemical reaction channel. 
The internal time $T_k(t)$ is defined as
\begin{equation}\label{eq:internal_time_definition}
	T_k(t)=\int_{0}^{t}a_k(X(s),s)ds
\end{equation}
And Eq. \ref{eq:space_system_evolve} is becoming
\begin{equation}\label{eq:system_evolve_internal_time}
	X(t) = X(0) + \sum_{k=1}^M Y_k\left(T_k(t)\right) v_k
\end{equation}
This is where the random time notion comes from. 

\section{Random time simulation algorithm}

In order to perform a complete stochastic simulation, two things must be cleared firstly,
\begin{enumerate}
	\item how much time passed before one of the stochastic processes,	$Y_k$, fires;
	\item which $Y_k$ fires at that later time;
\end{enumerate}
If we view the state of chemical reaction network as points in multi-dimensional phase space, the time evolution of this model can be viewed as random walked in spatial correlated non-homogeneous space. 
Then the two questions above is the fundamental questions about space and time. 

We will first determine the firing time problem. 
If we denote $Q(t, s)$ as the probability of no reaction occurred in the time interval $(s, t]$, then by the Markov property we have,
\begin{equation}\label{eq:Q_diff0}
	Q(t, s+\Delta s) = Q(t, s) Q(s, s+\Delta s)
\end{equation}
According to Eq. \ref{eq:conditional_probability}, the probability of no reaction occurred in the time interval $(s, s+\Delta s)$ is the prod of all independent reaction channels. 
\begin{equation}\label{eq:Q_diff1}
	Q(s, s+\Delta s) = \prod_{k=1}^M \left(1 - a_k(X(s), s) \Delta s\right)
\end{equation}
Therefore, 
\begin{equation}\label{eq:Q_diff2}
	\begin{aligned}
	Q(t, s+\Delta s) &= Q(t, s) \prod_{k=1}^M \left(1 - a_k(X(s), s) \Delta s \right) \\
	&=Q(t, s) \left(1 - \sum_{k=1}^M a_k(X(s), s) \Delta s + O((\Delta s)^2)\right) \\
	\end{aligned}
\end{equation}
By taking the limit of $\Delta s \rightarrow 0$, we arrive at the ordinary differential equation,
\begin{equation}\label{eq:Q_differential_equation}
	\frac{dQ(t, s)}{ds} = - Q(t, s) \sum_{k=1}^M a_k(X(s), s)
\end{equation}
Integral in the time interval $(t, s]$, and notice $a_k$ is not changed in the time interval $(t, s]$, we have
\begin{equation}\label{eq:Q_solution}
	Q(t, s) = \exp \left(- \sum_{k=1}^M a_k(X(t), t) (s - t)\right)
\end{equation}
Eq. \ref{eq:Q_solution} tells us if a random number $r$ is uniformly distributed on $[0, 1]$, then the time interval is
\begin{equation}\label{eq:time_interval}
	\Delta t = \ln (1 / r)
\end{equation}
In the random time representation of the stochastic model of chemical reaction network, for every infinitesimal time frame the stochastic process can be described by a counting process and further be decomposed into unit rate Poisson processes. 
All these unit rate Poisson processes are independent and remain stationary until some chemical reaction channel is fired. 
For $k$-th channel, follow the same argument from Eq. \ref{eq:Q_diff0} to Eq. \ref{eq:Q_differential_equation}, the internal time has the same distribution formula as Eq. \ref{eq:time_interval}. 
\begin{equation}\label{eq:internal_time_interval}
	\Delta T_k(t) = \ln \left(1/r_k\right)
\end{equation}
where $r_k$ is a random number uniformly distributed on $[0, 1]$. 
So if a set of random number is given $\left\{r_1, r_2, \ldots, r_M \right\}$, the corresponding real time for $k$-th chemical channel can be calculated from Eq. \ref{eq:internal_time_definition}. 
\begin{equation}\label{eq:calculate_internal_time}
	\Delta T_k(t) = \int_t^{t+\Delta t_k} a_k(X(s), s) ds
\end{equation}
Since all the decomposed processes are unit rate Poisson processes, event is expected to arrive in $\Delta T_k(t)$ for $k$-th chemical reaction channel. 
The real time passed when $k$-th chemical reaction channel arrived is $\Delta t_k$.
Accordingly the first fired chemical channel is the one that has the smallest internal time.
First arriving wins all. 
And the real time interval between two consequence reactions is
\begin{equation}\label{eq:smallest_internal_time}
\Delta t=min_k\{\Delta t_k\}
\end{equation}
Of course the firing chemical reaction channel is $k$-th channel.
Therefore we have answered the two main questions for stochastic simulations of the chemical reaction network using random time representation. 

\section{Next reaction random time algorithm}
According to the discussions in the former section, the simulation procedure by applying the random time algorithm can be listed as Table \ref{tab:first_simulation_procedure}.
\begin{table*}[!b]
	\centering
  \caption{Simulation procedure of random time algorithm}\label{tab:first_simulation_procedure}
	\begin{tabular}{|l|}
  \hline
  initialize time $t = 0$ with state $X(0) = \{X_1(0), X_2(0), \ldots, X_N(0)\}$ \\
	while $(t < t_{max})$ do: \\
	\ \ \ \ \ \ \ \ \ \ \  generate $M$ random numbers $r_k \sim U(0, 1)$, where $k = 1, \ldots, M$ \\
	\ \ \ \ \ \ \ \ \ \ \  set internal time interval $\Delta T_k = \ln(1/r_k )$ for all $k = 1, \ldots, M$ \\
	\ \ \ \ \ \ \ \ \ \ \  compute $\Delta t_k$ for all $k = 1,\ldots, M$ by solving $\int_{t}^{t+\Delta t_k} a_k(X(s),s)ds=\Delta T_k$ \\
	\ \ \ \ \ \ \ \ \ \ \  select reaction having $\Delta t=min_k\{\Delta t_k\}$, assume $j$-th channel is triggered\\
	\ \ \ \ \ \ \ \ \ \ \   update time $t=t+\Delta t$ \\
	\ \ \ \ \ \ \ \ \ \ \   update state $X(t+\Delta t)=X(t)+ v_j$ \\
  end while \\
	\hline
	\end{tabular}
\end{table*}

One thing about this simulation algorithm is it demands too many random numbers. 
For every time step, it needs $M$ random numbers to determine the time interval and firing reaction channel. 
Usually a stochastic simulation must be running for a long time to collect sufficient data points for statistical analysis. 
And the random number generators currently developed can only give a pseudorandom number sequence which will fail more likely in long sequences. 
The stochastic simulation algorithm that developed by Gillespie \cite{Gillespie1992} only demands two random numbers in each time step. 
If the next reaction algorithm is used, only one random number is demanded in one simulation loop. 

However, it is possible to develop an algorithm that only need one random number for each simulation step. 
It can be noticed the only random number that is consumed in one simulation step is the one that trigger the firing of a specific reaction channel. 
Other random numbers are untouched. 
This means the random numbers that do not trigger reaction channel firing can be reused. 
As a matter of fact, this discussion suggests that the internal time of unfired reaction channel is not changed after the end of this time frame. 
That is to say, if the triggered reaction channel is the $j$-th channel, we can continue the simulation by only refresh one internal time. 
\begin{equation}\label{eq:random_number_refresh}
	\begin{aligned}
	&\Delta T_j = \ln \left(1 / r\right)  \\
	&\Delta T_{k\neq j} =\Delta T_{k\neq j} \\
	\end{aligned}
\end{equation}
where $r$ is a random number uniformly distributed in $[0, 1]$.

This random time reuse algorithm is much like the Next Reaction Algorithm of stochastic simulation, so it is called the Next Reaction Random Time Algorithm (NRRTA). 
The simulation procedure of NRRTA is shown as Table \ref{tab:NRRTA_simulation_procedure}.
\begin{table*}[!h]
	\centering
  \caption{Simulation procedure of Next Reaction Random Time Algorithm}\label{tab:NRRTA_simulation_procedure}
	\begin{tabular}{|l|}
		\hline
		initialize time $t = 0$ with state $X(0) = \{X_1(0), X_2(0), \ldots, X_N(0)\}$ \\
		generate $M$ random numbers $r_k \sim U(0, 1)$ \\
		set internal time intervals $\Delta T_k = \ln(1/r_k )$ for all $k = 1, \ldots, M$ \\
		while $(t < t_{max})$ do: \\
		\ \ \ \ \ \ \ \ \ \ \  compute $\Delta t_k$ for all $k = 1,\ldots, M$ by solving $\int_{t}^{t+\Delta t_k} a_k(X(s),s)ds=\Delta T_k$ \\
		\ \ \ \ \ \ \ \ \ \ \  select reaction having $\Delta t=min_k\{\Delta t_k\}$, assume $j$-th channel is triggered \\
		\ \ \ \ \ \ \ \ \ \ \   update time $t=t+\Delta t$ \\
		\ \ \ \ \ \ \ \ \ \ \   update state $X(t+\Delta t)=X(t)+v_j$ \\
		\ \ \ \ \ \ \ \ \ \ \   generate a random number $r \sim U(0, 1)$ \\
		\ \ \ \ \ \ \ \ \ \ \   refresh internal time interval: $\Delta T_j = \ln \left(1 / r\right)$, $\Delta T_{k\neq j} =\Delta T_{k\neq j}$ \\
		end while \\
		\hline
	\end{tabular}
	\begin{flushleft}
		\small Notes: the boundary conditions can be achieved by restrict the firing channels. 
		Those channels that are forbidden for firing is assigned a frozen internal time.
	\end{flushleft}
\end{table*} 

\section{Implementations}

\begin{table*}[!h]
	\centering
	\caption{Parallel next reaction random time algorithm.}\label{tab:parallel-NRRTA}
	\begin{tabular}{|l|}
    \hline
    Initialize. Compute $a_k$, set $\Delta t_k = ln(1 / r_k)/a_k$, set $p_k=a_k$ \\
    while $(t < t_{max})$ do: \\
    \ \ \ \ \ \ \ \ \ \ \  Scan for $\Delta t_\mu = min_k\{\Delta t_k\}$ \\
    \ \ \ \ \ \ \ \ \ \ \  Update species for reaction $\mu$, set $t = t + \Delta t_\mu$ \\
    \ \ \ \ \ \ \ \ \ \ \  Recalculate $a_k$ for every reaction, set $\Delta t_\mu = ln(1 / r)/a_\mu$, set $\Delta t_k = p_k/a_k(\Delta t_k - \Delta t_\mu)$ \\
		end while \\
		\hline
	\end{tabular}
\end{table*} 

\begin{table*}[!h]
	\centering
	\caption{Model parser.}\label{tab:model-parser}
	\begin{tabular}{|l|}
    \hline
    // simple model of Lotka predator oscillation \\
      \# init \\
      A = 1000 \\
      B = 1000 \\
      \# reactions \\
      1 A -\ -$>$ 2 A : 10 \\
      1 A 1 B -\ -$>$ 2 B : 0.01 \\
      1 B -\ -$>$ NULL : 10 \\
      \# end \\
      time = 2 \\
      steps = 10000 \\
		\hline
	\end{tabular}
\end{table*} 

The most crucial and time consuming step with next reaction random time algorithm is the selection of the minimum inner time $\Delta t_\mu$. For memory efficiency and algorithm simplicity, the algorithm was modified as outlined in Table~\ref{tab:parallel-NRRTA}.

The most accepted model file formate for current system biology is the SBML (System Biology Markup Language). However, there is a lack of collective graphic view of the model file, making the edition of these files not convinient for ordinary use. We used a different model file formate which can give both insight of the while network and also provide all the informations for GSSA simulations. A typical predator oscillation system can be expressed as in Table~\ref{tab:model-parser}.

Since biology network follow the scale-free (or power-law) topography, the reactant matrix, stochiometric matrix and species update matrix are intrinsically sparse. Memory consuming is a potential issue for handling large network which could consist of over $10^2$ nodes, in other words, over $10^4$ reactions. Sparse matrix is the natural choice for biological network simulation. For achieving $O(1)$ accessibiligy, we chose the COO (coordinate formated matrix) formate for sparse matrix storage and computation.

In a large network, sequentially update of all the reactions after a firing process is not a wise idea. Therefore the denpendency graph was proposed for the efficient update of the network state. The idea was first published by (Michael A. Gibson and Jehoshua Bruck, 2000, JPC). By following the network connection through the affected species, a dependency graph can be constructed as illustrated in the following graph. 
However, for massive parallel computation, the network state is updated concurrently, so no dependency graph is used in the cuda algorithm.

The key property for cuda programming is the memory arrangement, since memory access latency is about 20-30 times longer than arithmetic operations. Although memory access can be hidden partially by parallel execution of warps by SMs, cooperate with the GPU memory caches is still vital.
The utilizing of registers and shared memory contribute to more efficient latency hidding for parallel algorithm. For detail of this part, refer to (Professional CUDA C Programming, 2014).
In the original cuda programming set, computing hierarchy is arranged as the grid-block 2 layer structure. While for natural setup for parallel simulation of biology network, an intermediate layer called the "bundle" is used for holding an independent trajectory. The reason for the need of this intermediate layer is the limiting of the threads that could reside in a particular block. Also for best performance, the threads per block should better be $2^n$ for efficient reduction. Given this considerations, the simulation architecture is arranged as the following. 
\begin{itemize}{}{}
  \item 1 trajectory corresponds to 1 bundle
  \item bundle consists of a batch of blocks
  \item 1 thread in a block corresponds to 1 reaction related computation
\end{itemize}

The best pesudo-random number generator we discover is the Mersenne Twister (MT) algorithm. With proper parameter values, MT can generate sequence with a period as long as $2^{19,937}$ and extremely good statistical properties. However, when implement MT in GPU architecture, the algorithm operation speed was limited by the bandwidth of memory. For solving this issue, we had implemented an modified version of the GPU MT algorithm. The generation speed was increased but due to the limit of the size of shared memory, maximum 256 threads were allowed in a single block, and totally 256 blocks were allowed in a single run. There are possibilities that the implementation can be improved, but that involved with complete rewrite the algorithm in a self-made version which demand much effort and time. So we changed the idea of a pure GPU implementation to a heterogeneous version. The CPU were responsible for the generation of random numbers and then passed to the global memory of GPU. This implementation works fine for small trajectory number and short simulation time. But when simulation trajectory number is large or simulation time is very long, the global memory would be exhuasted. The situation can be solved with different streams of GPU computation. Depending on the computation latency of simulation, several streams can be performed to simulate different trajectories while at the same time performing memory copy from CPU memory heap to GPU global memory. 

We find shared memory are vital for GPU implementation in various ways. One problem of using shared memory is the memory size limitation. By using sparse representation and limiting the biology network to maximum 2 reactants, matrices were able to be fitted into the shared memory of each block. In most biology network, the reactants is less than 2. Even for reactions involved with 3 reactants, the reaction can also be transformed into an equivalence 2 reactants formate. With this setup, the loading from global memory can be performed coasleced, also bank conflicing is avoided to some extent. 

The most time consuming step in the GPU GSSA algorithm is the reduction step for finding the next reaction id and fire time. For classic reduction algorithm, the algorithm were optimized for $2^n$ elements reduction. For biology network, the reaction number can be arbitrarily valued, therefore the classic reduction algorithm should be modified to efficiently handle arbitrary number of element. Another complexity is the reduction across different blocks. Blocks within a bundle should be reduced together, while blocks belong to different bundles should not be reduced. Therefore a two level reduction is needed. With these considerations in mind, we developed an optimized version for block reduction and this kernel can also be applied to bundle reduction as well. This part forms the loop unroll template function which is the core part of the reduction computation. 

\begin{figure}[t]
    \begin{center}
    \includegraphics[scale=0.6]{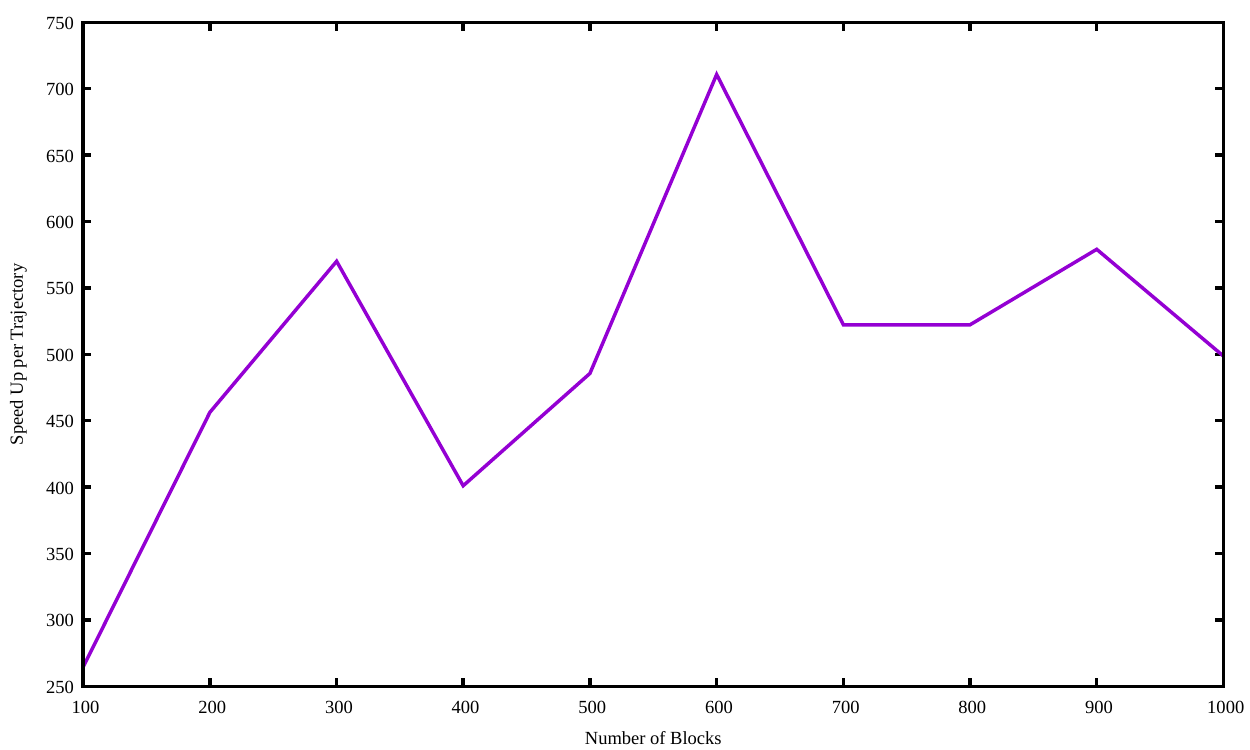}
    \end{center}
    \caption{
      \textbf{Parallel next reaction random time stochastic simulation speedup of predator-prey model.}
      Speed up of parallel version of next reaction random time stochastic simulation algorithm with predator-prey model. 
      The computation was done using a graphic card of Nvidia RTX 1060 with capability 6.1, global memory size 6GB. 
      10000 trajectories were run simultaneously, with 100000 steps for each trajectory.}
    \label{fig:speedup}
\end{figure}

The performance of the GPU version of GSSA depends on the capability of the graphic card, also on the hierarchy computation units of cuda kernel. In our current setup, when using Nvidia RTX 1060 with capability 6.1, global memory size 6GB, about 400-500 folds speed-up can be achieved with single-precision floating point number as shown in Figure~\ref{fig:speedup}. The speed up can be expected to further increase when using a more powerful GPU.

\section{Discussions}
The major time consuming step for Gillespie's stochastic simulation algorithm is the calculation of the time interval between two consecutive reactions as well as the searching for firing reaction channel. 
The searching step in random time simulation algorithm is solved by finding the reaction channel that has the minimal expected real time. 
However the time consumed by calculating the time interval is not much accelerated by the random time simulation algorithm as compared with Gillespie's simulation algorithm. 
For a time-inhomogeneous chemical reaction system the time interval for Gillespie's stochastic algorithm can be calculated from Eq. \ref{eq:propensity_function_definition} and Eq. \ref{eq:time_interval}. 
\begin{equation}\label{eq:Gillespie_calculate_time}
	\ln \left(1/r\right) =\sum_{k=1}^M \int_t^{t+\Delta t} c_k(s) \prod_j \frac{X_k(t)!}{\sigma_{jk}!(X_k(t) - \sigma_{jk})!} ds
\end{equation}
While for random time stochastic simulation algorithm, the time interval can be obtained by solving the following integral equation group. 
\begin{equation}\label{eq:random_time_calculate_time}
	\begin{aligned}
  &\ln \left(1/r_k\right) =\prod_j \frac{X_k(t)!}{\sigma_{jk}!(X_k(t) - \sigma_{jk})!} \int_t^{t+\Delta t_k} c_k(s)ds,\\
  & k = 1, \ldots, M \\
	&\Delta t = min_k\left\{\Delta t_k\right\} \\
	\end{aligned}
\end{equation}
For large-scale chemical reaction network, the only hope for solving Eq. \ref{eq:Gillespie_calculate_time} is through numerical integration. 
While for the random time simulation algorithm, the time evolution is separated from the summation of propensity functions, so integral can be calculated analytically for well behaved time-dependent reaction rates. 

If we assume the number of molecules in the system is on order of $N$, the number of reaction channel is on order of $M$, the stoichiometric number is on order of $H$, the number of reaction species for one single reaction is on order of $Q$ and the integral steps is on order of $I$. 
In order to find $\Delta t$, we apply binary tree searching to find the next reaction, the element operation number is on order of $R$. 
So the number of element operations for calculate the propensity function is on order of $2NQ+3Q$. 
And the number of element operations for solving Eq. \ref{eq:Gillespie_calculate_time} is on order of $\left[MI(2NQ+3Q)+M-1\right]R$. 
For Eq. \ref{eq:random_time_calculate_time}, with well behaved time-dependent perturbubation, the integral can be solved analytically, so the calculation is reduced to a single element operation. 
The number of element operations for solving Eq. \ref{eq:random_time_calculate_time} is on order of $[2NQ+3Q+2]M$. 
For large-scale chemical reaction network, like $10^4$ chemical species, $M$ is on the order of $10^8$. 
While if we consider a real biology chemical reaction network, $N$ is typically on the order of $10^3$. 
The stoichiometric number for a biology system is typically less than $10$, and reaction species involved in one biochemical reaction is also less than $10$. 
The integral steps is related to the calculation accuracy of the results since $I = \Delta t / \delta s$, here we assume it to be on order of $10^4$. 
So the number of element operations for solving Eq. \ref{eq:Gillespie_calculate_time} is on order of $10^{16}$, while for Eq. \ref{eq:random_time_calculate_time} the number is reduced to an order of $10^{10}$. 

But the accelerations can not be the same of computed above.
The main reason for not achieving the high speed include the memory limitation of the GPU hardware, the limit number of the computation core and also the limitation of total thread numbers. 
On the other side, for Eq. \ref{eq:Gillespie_calculate_time}, the integration of propensity function can be made parallel. 
In this circumstances, the summation of contributions of propensity from all the reaction channels demands a reduction algorithm, which followed by a searching process that locate the next reaction. 
However, the situation is very different in solving Eq. \ref{eq:random_time_calculate_time}. 
Since no summation is demanded, the summation and searching are done within one single reduction process. 
This means at the most time consuming step, parallel random time algorithm can be approximate 2 times faster than parallel Gillespie's stochastic simulation algorithm. 
Hence random time based algorithms are more appreciate for parallel computations. 

\section{ACKNOWLEDGEMENTS}

Chuanbo Liu thanks supports by Natural Science Foundation of China, No.32000888,
Jin Wang thanks the supports from grant no. NSF-PHY 76066 and NSF-CHE-1808474,
Ministry of Science and Technology of China, No.2016YFA0203200, 
Projects of Science and Technology Development, Jilin, China, No.20180414005GH,
Projects of Instrument and Equipment Development, Chinese Academy of Sciences, No.Y928041001. 












\end{document}